\documentclass[preprint,showpacs,preprintnumbers,amsmath,amssymb]{revtex4}

\usepackage{graphicx}
\usepackage{dcolumn}
\usepackage{bm}
\def\E^#1{{\buildrel #1 \over\vee}}

\begin{document}

\vphantom{math}
\vskip1.0cm

\title{FROM BBGKY HIERARCHY TO NON-MARKOVIAN EVOLUTION EQUATIONS}%

\author{V.I. GERASIMENKO, V.O. SHTYK$^\dag$, A.G. ZAGORODNY$^\dag$}%

\email{gerasym@imath.kiev.ua,
 vshtyk@bitp.kiev.ua, azagorodny@bitp.kiev.ua}

\affiliation{Institute of Mathematics NAS of Ukraine\\
3, Tereshchenkivs'ka str., Kyiv 01601\\
$^\dag$Bogolyubov Institute for Theoretical Physics NAS of Ukraine\\
14-b, Metrolohichna str., Kyiv, 03680}%

\begin{abstract}
The problem of description of the evolution of the microscopic phase density and its generalizations is discussed.
With this purpose, the sequence of marginal microscopic phase densities is introduced, and the appropriate BBGKY
hierarchy for these microscopic distributions and their average values is formulated.
The microscopic derivation of the generalized evolution equation for average value of the microscopic phase density
is given and the non-Markovian generalization of the Fokker-Planck collision integral is deduced.
\end{abstract}


\maketitle
\tableofcontents

\newpage
\vphantom{math}
\vskip2.5cm

\section{Introduction}
Recently, much interest has been generated to the studies of physical systems with
statistical properties which cannot be described within the concept of the Markovian random processes.
This concerns anomalous transport in turbulent plasma, strange diffusion of magnetic field lines in fusion plasma,
the Brownian motion of macroparticles in complex fluids \emph{etc}. (see, for example, \cite{Ba95,Ba97,KL,A,MK,SK,Ba05,SZ}
and references cited therein). One of the problems challenging and most important
at the same time is to describe the kinetic properties of such systems and to understand the role of non-Markovian
effects of particle and energy transport in such systems. The key point of such calculations is the formulation of the consistent kinetic-type
equations with regard to time-nonlocality of the collision integrals and the renormalization of the free-streaming
particle propagator due to fluctuation influence on particle trajectory. In \cite{ZW99,ZW01,ZW09}, this problem
was solved on the basis of the Green function method similar to that proposed in \cite{OK}. The main disadvantage
of such a treatment is the uncertainty introduced by the specific splitting of higher correlations made on the basis
of some physical arguments.

The aim of the present paper is to derive the non-Markovian kinetic-type equations in a more consistent way on the basis
of a rigorous solution of the appropriate BBGKY hierarchy (Bogolyubov-Born-Green-Kirkwood-Yvon hierarchy) for the generalized
microscopic phase densities and their average values.

The paper is organized in the following order.

In Section II, we introduce the definitions used for the
description of the evolution of
the microscopic phase density and its generalization. In Section III, we deduce the evolution equations
for average values of microscopic phase densities and
construct a solution of initial-value problem of the obtained hierarchy of equations.
In Section IV, with the use of the results obtained above, we develop a new approach to the description of the evolution of the
average value of microscopic phase density; namely, we construct an evolution equation for
this quantity and deduce a
non-Markovian generalization of the Fokker-Planck collision integral.


\section{The evolution of the marginal microscopic phase\\ densities}
We consider the system of a non-fixed
(i.e. arbitrary but finite) number of identical particles with unit mass $m=1$ in the space $\mathbb{R}^3$
(in the terminology of statistical mechanics it is known as nonequilibrium grand canonical ensemble \cite{CGP97}).
Every particle is characterized by the phase space coordinates  $x_i\equiv(q_i,p_i)$,
i.e. by a position in the space $q_i\in \mathbb{R}^3$ and a momentum $p_i\in \mathbb{R}^3$.
A description of many-particle systems is formulated in
terms of two sets of objects: by the sequence of observables
and by the sequence of states \cite{Ger4}.

\subsection{The dual BBGKY hierarchy}
In \cite{Ger4} it was introduced the marginal observables known as the microscopic phase densities
$G(t)=\big(G^{(1)}(t),\ldots,$ $ G^{(k)}(t),\ldots\big)$  of $k$-ary type
$G^{(k)}(t)=\big(0,\ldots,0,G_{k}^{(k)}(t),\ldots,$ $G^{(k)}_{s}(t),\ldots\big)$.

For example, at the initial time moment $t=0$ the sequence of marginal
additive-type microscopic phase densities \cite{Ger4} has the form
$G^{(1)}(0)=\big(0,\ldots,\delta(\xi_{1}-x_1),0,\ldots\big)$, where $\delta$ is the Dirac $\delta$-function.
Correspondingly, the sequence of marginal observables
of $k$-ary type, $k\geq1$, microscopic phase densities (\ref{k-arN}) is given as follows:
\begin{eqnarray}\label{ad_G0}
  && G^{(k)}(0)=\big(0,\ldots,0,\sum\limits_{i_1\neq\ldots\neq i_k=1}^{k}\prod\limits_{l=1}^{k}\delta(\xi_l-x_{i_l}),0,\ldots\big).
\end{eqnarray}

Let $H_n=\sum\limits{_{i=1}^{n}}\frac{p_{i}^{2}}{2}+
  \sum\limits{_{i<j=1}^{n}} \Phi(q_{i}-q_{j})$
is the Hamiltonian of a $n$-particle system, $\Phi(q_{i}-q_{j})$ is a two-body interaction potential,
$Y\equiv(x_1,\ldots,x_s)$,
$(x_1,\ldots,\E^{j},\ldots,x_s)\equiv(x_1,\ldots,x_{j-1},x_{j+1},\ldots,x_s)= Y\setminus x_{j}$ 
and the brackets $\langle \cdot,\cdot\rangle$ is a scalar product of vectors.

The marginal microscopic phase densities
$G^{(k)}_{s}(t)\equiv G^{(k)}_{s}(t,\xi_1,\ldots,\xi_k;x_1,\ldots,x_s)$  of every $k$-ary type
are governed by the initial-value problem of the dual BBGKY hierarchy \cite{Ger4},\cite{BG}
\begin{eqnarray} \label{dual1}
    &&\frac{\partial}{\partial t}G^{(k)}_{s}(t)=
    \big(\sum\limits_{i=1}^{s}\langle\, p_i,\frac{\partial}{\partial q_i}\rangle-
    \sum\limits_{i\neq j=1}^{s}\langle\frac{\partial}{\partial q_i}\Phi(q_i-q_j),\frac{\partial}{\partial p_i}\rangle\big) G^{(k)}_{s}(t)\nonumber\\
&&-\sum\limits_{i\neq j=1}^{s}\langle\frac{\partial}{\partial q_i}\Phi(q_i-q_j),
      \frac{\partial}{\partial p_i}\rangle G^{(k)}_{s-1}(t,Y\setminus x_{j})
\end{eqnarray}
with the initial data
\begin{eqnarray} \label{dual2}
        && G^{(k)}_{s}(t)\mid_{t=0}=G^{(k)}_{s}(0),\quad s\geq k\geq 1.
\end{eqnarray}

As a case in point, we adduce the first equation of hierarchy (\ref{dual1})
\begin{eqnarray*}
   && \frac{\partial}{\partial t}G^{(k)}_{k}(t)=
    \big(\sum\limits_{i=1}^{k}\langle\, p_i,\frac{\partial}{\partial q_i}\rangle-
\sum\limits_{i\neq j=1}^{k}\langle\frac{\partial}{\partial q_i}\Phi(q_i-q_j),
      \frac{\partial}{\partial p_i}\rangle\big) G^{(k)}_{k}(t).
\end{eqnarray*}
For the marginal additive-type microscopic phase density, the first two equations have the form
\begin{eqnarray*}
   && \frac{\partial}{\partial t}G^{(1)}_{1}(t,\xi_1;x_1)=
    \langle\, p_1,\frac{\partial}{\partial q_1}\rangle G^{(1)}_{1}(t,\xi_1;x_1),\\
   && \frac{\partial}{\partial t}G^{(1)}_{2}(t,\xi_1;x_1,x_2)=
    \big(\sum\limits_{i=1}^{2}\langle\, p_i,\frac{\partial}{\partial q_i}\rangle-\sum\limits_{i\neq j=1}^{2}\langle\frac{\partial}{\partial q_i}\Phi(q_i-q_j),\frac{\partial}{\partial p_i}\rangle \big) G^{(1)}_{2}(t,\xi_1;x_1,x_2)-\\
    &&-\sum\limits_{i\neq j=1}^{2}\langle\frac{\partial}{\partial q_i}\Phi(q_i-q_j),
      \frac{\partial}{\partial p_i}\rangle G^{(1)}_{1}(t,\xi_1;x_{i}).
\end{eqnarray*}


Let
$N^{(k)}(t)=\big(0,\ldots,0,$ $N_{k}^{(k)}(t),\ldots,N^{(k)}_{n}(t),\ldots\big)$,  $k\geq 1$, is a sequence 
of microscopic phase densities of $k$-ary type \cite{SZ}
\begin{eqnarray}\label{k-arN}
  &&N_{n}^{(k)}(t)\equiv N_{n}^{(k)}(t,\xi_1,\ldots,\xi_k;x_1,\ldots,x_n)=\sum\limits_{i_1\neq\ldots\neq i_k=1}^{n}\prod\limits_{l=1}^{k}\delta(\xi_{l}-X_{i_l}(t,x_1,\ldots,x_n)),
\end{eqnarray}
where $\xi_1,\ldots,\xi_k$ are the macroscopic variables
$\xi_i=(v_i,r_i)\in\mathbb{R}^3\times\mathbb{R}^3$ and the set of functions
$\big\{X_{i}(t,x_1,\ldots,x_n)\big\}_{i=1}^{n}$, $n\geq k\geq 1$, are the solution
of the Cauchy problem of the Hamilton equations for $n$ particles with the initial data $x_1,\ldots,x_n$.
Then the sequence of marginal observables $G(t)$ in terms of sequence (\ref{k-arN}) are defined by the formula \cite{Ger4}
\begin{eqnarray}\label{mo}
   &&G_{s}^{(k)}(t,x_1,\ldots,x_s)=\sum_{n=0}^s\,\frac{(-1)^n}{n!}\sum_{j_1\neq\ldots\neq j_{n}=1}^s
   N_{s-n}^{(k)}\big(t,\xi_1,\ldots,\xi_k;Y\backslash \{x_{j_1},\ldots,x_{j_{n}}\}\big),
\end{eqnarray}
where $Y\equiv(x_1,\ldots,x_s)$,\, $s\geq 1$.

For example, if $k=1$, i.e. in the case of an additive-type observable,
we have the microscopic phase density \cite{SZ,ZW99}
\begin{eqnarray*}
   &&N_{n}^{(1)}(t,\xi_1;x_1,\ldots,x_n)=\sum\limits_{i=1}^{n}\delta(\xi_{1}-X_{i}(t,x_1,\ldots,x_n)).
\end{eqnarray*}

We note that expression (\ref{k-arN}) defines
the one-parametric group of operators $\mathbb{R}^1\ni t\mapsto S_n(t)N_{n}(0)$, i.e.
\begin{eqnarray}\label{sn}
    &&N_{n}^{(k)}(t,\xi_1,\ldots,\xi_k;x_1,\ldots,x_n)= S_n(t)N_{n}^{(k)}(0),
\end{eqnarray}
where $N_{n}^{(k)}(0)$ is the initial microscopic phase density (\ref{k-arN}).


Let $Y=(x_1,\ldots,x_s),$\, $X=Y\setminus\{x_{j_1},\ldots,x_{j_{s-n}}\}$.
For continuous functions in the capacity of initial data, a solution of Cauchy problem (\ref{dual1})-(\ref{dual2})
is defined by the following expansion
\begin{eqnarray}\label{rozvG}
   && G^{(k)}_s(t,Y)=\sum\limits_{n=0}^{s}\frac{1}{n!}\sum\limits_{j_1\neq\ldots\neq j_n=1}^{s}\mathfrak{A}_{1+n}(t,\big(Y\setminus X\big)_1,X)
    G^{(k)}_{s-n}(0,Y\setminus \{x_{j_1},\ldots,x_{j_n}\}),
\end{eqnarray}
where the evolution operator
\begin{eqnarray*}
   &&\mathfrak{A}_{1+n}(t,\big(Y\setminus X\big)_1,X)=
  \sum\limits_{\mathrm{P}:\{(Y\setminus X)_1,X\} ={\bigcup\limits}_i X_i}(-1)^{|\mathrm{P}|-1}(|\mathrm{P}|-1)!
   \prod_{X_i\subset \mathrm{P}}S_{|X_i|}(t,X_i)
\end{eqnarray*}
is the $(1+n)th$-order cumulant of groups $S_{|X_i|}(t)$
of operators (\ref{sn}), ${\sum\limits}_\mathrm{P}$
is the sum over all possible partitions $\mathrm{P}$ of the set $\big\{(Y\setminus X)_1,X\big\}$ into
$|\mathrm{P}|$ nonempty mutually disjoint subsets $ X_i\subset \big\{(Y\setminus X)\big)_1,X\big\}$.
The set $(Y\setminus X)_1$ consists of one element of $Y\backslash X$,
i.e. the set $Y\backslash X=\{x_{j_1},\ldots,x_{j_{s-n}}\}$
is a connected subset of the partition $\mathrm{P}$ ($|\mathrm{P}|=1$).

Then in terms of variables $\xi_1,\ldots,\xi_k$, for the marginal microscopic phase densities of
$k$-ary type $G^{(k)}(t)=\big(0,\ldots,0,G_{k}^{(k)}(t),\ldots,G^{(k)}_{s}(t),\ldots\big)$
we derive
\begin{eqnarray} \label{k_lanc}
  &&\frac{\partial}{\partial t}G^{(k)}_{s}(t)=\big(-\sum\limits_{i=1}^{k}\langle v_i,\frac{\partial}{\partial r_i}\rangle+
   \sum\limits_{i\neq j=1}^{k}\langle \frac{\partial}{\partial r_i}\Phi(r_i-r_j),
   \frac{\partial}{\partial v_i}\rangle \big)G^{(k)}_{s}(t)+\nonumber\\
  &&   +\sum\limits_{i=1}^{k}\int d\xi_{k+1}\langle\frac{\partial}{\partial r_i}\Phi(r_i-r_{k+1}),
     \frac{\partial}{\partial v_i}\rangle G^{(k+1)}_{s}(t)
\end{eqnarray}
with the initial data
\begin{eqnarray} \label{k_lancin}
    &&G^{(k)}_{s}(t)\mid_{t=0}=\sum\limits_{i_1\neq\ldots\neq i_k=1}^{s}
    \prod\limits_{l=1}^{k}\delta(\xi_l-x_{i_l})\delta_{s,k}.
\end{eqnarray}
Here, $1\leq r<s$, and if $k=s$, the marginal microscopic phase density $G^{(s)}_{s}(t)$ is governed by the Liouville equation.

Thus, in terms of variables $\xi_1,\ldots,\xi_k$,
the dual BBGKY hierarchy (\ref{k_lanc}) for marginal microscopic phase densities (\ref{rozvG})
is represented as the Bogolyubov equations set with respect
the arity index $k\geq1$,
while evolution equations (\ref{k_lanc}) have a structure of a sequence of equations
with respect to the index of the number of particles $s\geq k$.

We note that, for every $s$, a solution of initial-value problem (\ref{k_lanc})-(\ref{k_lancin}) of
equations set (\ref{k_lanc}) can be represented as the following expansion:
\begin{eqnarray*}
 && G^{(k)}_{s}(t,\xi_1,\ldots,\xi_k;x_1,\ldots,x_s)=\\
&&=\sum\limits_{n=0}^{s-k}\,\frac{1}{n!}\int d\xi_{k+1}\ldots d\xi_{k+n}\,
      \mathfrak{A}_{1+n}(-t,Y_{1},\xi_{k+1},\ldots,\xi_{k+n})\,G^{(k+n)}_{s}(0),
\end{eqnarray*}
where $G^{(k+n)}_{s}(0)$ is initial data (\ref{k_lancin}), and
$\mathfrak{A}_{1+n}(-t,Y_{1},\xi_{k+1},\ldots,\xi_{k+n})$
is the $(1+n)th$-order cumulant (\ref{cm}) of the groups of evolution operators (\ref{eo}) defined further.

\subsection{On the evolution of states of many-particle systems}
We furnish further comments about the evolution of microscopic phase densities in the framework of the evolution of states
of many-particle systems.

In this case, the microscopic phase densities are defined at the initial time moment $N(0)=(N^{(1)}(0),\ldots,N^{(k)}(0),\ldots)$,
where $N^{(k)}(0)=(0,\ldots,0,$ $N_{k}^{(k)}(0),\ldots,N_{n}^{(k)}(0),\ldots)$ and
\begin{eqnarray*}
&&N_{n}^{(k)}(0)\equiv N_{n}^{(k)}(0,\xi_1,\ldots,\xi_k;x_1,\ldots,x_n)=
\sum\limits_{i_1\neq\ldots\neq i_k=1}^{n}\prod\limits_{l=1}^{k}\delta(\xi_{l}-x_{i_l}), \quad n\geq k\geq 1.
\end{eqnarray*}

The evolution of states is usually described by the Cauchy problem of a sequence of the Liouville equations for
the sequence $D(t)=\big(I,D_{1}(t),\ldots,D_{n}(t),\ldots\big)$ of distribution functions
 $D_{n}(t)\equiv D_{n}(t,x_1,\ldots,x_n)$
\begin{eqnarray}\label{z-ks1}
     &&\frac{\partial}{\partial t}D_{n}(t)=\big(-\sum\limits_{i=1}^{n}\langle p_i,\frac{\partial}{\partial q_i}\rangle+
  \sum\limits_{i\neq j=1}^{n}\langle \frac{\partial}{\partial q_i}\Phi(q_i-q_j),\frac{\partial}{\partial p_i}\rangle\big) D_{n}(t).
\end{eqnarray}
\begin{eqnarray}\label{z-ks2}
&&D_{n}(t)|_{t=0}=D_{n}(0),\quad n\geq1.
\end{eqnarray}

A solution of the Cauchy problem (\ref{z-ks1})-(\ref{z-ks2}) is constructed in \cite{CGP97}.
Average values of the microscopic phase densities $N(0)$ are determined by the expressions
\begin{eqnarray}\label{est}
  &&\langle N^{(k)} \rangle (t,\xi_1,\ldots,\xi_k)=\nonumber\\
  &&=\big(1,D(0)\big)^{-1}\sum\limits_{n=0}^{\infty}\frac{1}{n!}
       \int dx_{k+1}\ldots dx_{k+n}D_{k+n}(t,\xi_1,\ldots,\xi_k,x_{k+1},\ldots,x_{k+n}),
\end{eqnarray}
where $D_k(t,\xi_1,\ldots,\xi_k,x_{k+1},\ldots,x_{k+n})$ is the value of
a solution of initial-value problem (\ref{z-ks1})-(\ref{z-ks2})
at a point $\xi_1,\ldots,\xi_k,x_{k+1},\ldots,x_{k+n}$. Thus, the evolution of
the functions $\langle N^{(k)} \rangle (t),\, k\geq1,$ will be governed
by the BBGKY hierarchy with respect to the arity index $k$ (see next section).

In the thermodynamic limit, the evolution of states is described in terms of the marginal distribution functions
governed by the BBGKY hierarchy \cite{CGP97},\cite{BC}. From Liouville equations (\ref{z-ks1})
for the marginal distribution functions $F(t)=\big(I,F_{1}(t),\ldots,F_{s}(t),\ldots\big)$, we derive
\begin{eqnarray}\label{1}
       &&\frac{\partial}{\partial t}F_{s}(t)=\big(-\sum\limits_{i=1}^{s}\langle p_i,\frac{\partial}{\partial q_i}\rangle +
       \sum\limits_{i\neq j=1}^{s}\langle\frac{\partial}{\partial q_i}\Phi(q_i-q_j),
      \frac{\partial}{\partial p_i}\rangle \big) F_{s}(t)+\nonumber\\
      &&+\sum\limits_{i=1}^{s}\int dx_{s+1}\langle\frac{\partial}{\partial q_i}\Phi(q_i-q_{s+1}),
      \frac{\partial}{\partial p_i}\rangle F_{s+1}(t),
\end{eqnarray}
\begin{eqnarray}\label{2}
      &&  F_{s}(t)\mid_{t=0}=F_{s}(0),\quad s\geq 1.
\end{eqnarray}

A solution of Cauchy problem (\ref{1})-(\ref{2}) was constructed in \cite{CGP97},\cite{GerRS}. The microscopic phase densities
in this case are given by marginal microscopic phase densities (\ref{ad_G0}) and their average values
at time moment $t\in \mathbb{R}$
\begin{eqnarray}\label{avmarSch}
  &&\big\langle A \big\rangle(t)=\big(G(0),F(t)\big)=
  \sum\limits_{s=0}^{\infty}\frac{1}{s!}
  \int dx_{1}\ldots dx_s G_{s}(0)F_{s}(t)=\nonumber\\
  &&=\big(G(t),F(0)\big)=
   \sum\limits_{s=0}^{\infty}\frac{1}{s!}\int dx_{1}\ldots dx_s G_{s}(t)F_{s}(0).
\end{eqnarray}
are determined by the expressions
\begin{eqnarray*}
  \langle G^{(s)} \rangle (t)=F_s(t,\xi_1,\ldots,\xi_s),\quad s\geq1,
\end{eqnarray*}
where $F_s(t,\xi_1,\ldots,\xi_s)$ is the value of a solution of initial-value problem (\ref{1})-(\ref{2})
at a point $\xi_1,\ldots,\xi_s$.
The evolution equations for the average values of microscopic phase densities will be considered in the next section.


\section{Evolution of average values of microscopic phase\\ densities}
We derive the evolution equations of average values of microscopic phase densities and
construct a solution of the initial-value problem of the obtained hierarchy of equations.

\subsection{The BBGKY hierarchy for average values of the marginal microscopic\\ phase densities}
We consider the general case, i.e. for the $k$-ary type microscopic
phase density. According to (\ref{avmarSch}) and (\ref{rozvG}), from the dual BBGKY hierarchy (\ref{k_lanc}) we derive
\begin{eqnarray}\label{Gs1}
    &&\frac{\partial}{\partial t}\langle G^{(k)} \rangle (t)=-\sum\limits_{i=1}^{k}\big\langle v_i,
    \frac{\partial}{\partial r_i}\big\rangle \langle G^{(k)} \rangle (t)+
    \sum\limits_{i\neq j=1}^{k} \langle \frac{\partial}{\partial r_i}\Phi(r_i-r_j),
    \frac{\partial}{\partial v_i} \rangle \langle G^{(k)} \rangle (t)+\nonumber\\
    &&+\sum\limits_{i=1}^{k}\int d\xi_{k+1} \langle\frac{\partial}{\partial r_i}\Phi(r_i-r_{k+1}),
    \frac{\partial}{\partial v_i} \rangle \langle G^{(k+1)} \rangle (t),
\end{eqnarray}
with the initial data
\begin{eqnarray}\label{Gs2}
    &&\langle G^{(k)} \rangle (t,\xi_1,\ldots,\xi_k)|_{t=0}=\langle G^{(k)} \rangle(0), \quad k\geq1.
\end{eqnarray}
Due to functional (\ref{avmarSch}), initial data (\ref{Gs2}) are given as the functions
\begin{eqnarray*}
   \langle G^{(k)} \rangle(0,\xi_1,\ldots,\xi_k)=F_{k}(0,\xi_1,\ldots,\xi_k),
\end{eqnarray*}
where
$F_s(0,\xi_1,\ldots,\xi_s)$ is the value of the initial marginal state at a point $\xi_1,\ldots,\xi_s$.

We note that, according to the definition of functionals (\ref{est}) and (\ref{avmarSch}),
 the equality
$ \langle G^{(k)} \rangle (t)=\langle N^{(k)} \rangle (t)$ holds in the case of finitely many particles.
In the thermodynamic, limit the value $\langle N^{(k)} \rangle (t)$ tends to $\langle G^{(k)} \rangle (t)$,
i.e. to the solution of Cauchy problem (\ref{Gs1})-(\ref{Gs2}).

\subsection{The BBGKY hierarchy for average values of the microscopic phase densities}
For comparison, we derive hierarchy (\ref{Gs1}) from equations (\ref{z-ks1}).
In the general case, i.e. the $k$-ary type microscopic
phase density (\ref{k-arN}), we have
\begin{eqnarray}\label{z-Bs1}
   &&\frac{\partial}{\partial t}\langle N^{(k)} \rangle (t)=
   \big(-\sum\limits_{i=1}^{k}\langle v_i,\frac{\partial}{\partial r_i}\rangle+
   \sum\limits_{i\neq j=1}^{k}\langle \frac{\partial}{\partial r_i}\Phi(r_i-r_j),
  \frac{\partial}{\partial v_i}\rangle \big)\langle N^{(k)} \rangle (t)+\nonumber\\
&&+\sum\limits_{i=1}^{k}\int d\xi_{k+1}\langle\frac{\partial}{\partial r_i}\Phi(r_i-r_{k+1}),
       \frac{\partial}{\partial v_i}\rangle \langle N^{(k+1)} \rangle (t),
\end{eqnarray}
with the initial data ($ k\geq1$)
\begin{eqnarray}\label{Bs2}
  && \langle N^{(k)} \rangle (t,\xi_1,\ldots,\xi_k)|_{t=0}=\big(1,D(0)\big)^{-1}\sum\limits_{n=0}^{\infty}\frac{1}{n!} \int dx_{k+1}\ldots dx_{k+n}D_{k+n}(0),
\end{eqnarray}
where $D_{k+n}(0)\equiv D_{k+n}(0,\xi_1,\ldots,\xi_k;x_{k+1},\ldots,x_{k+n})$ is the value of the distribution function
$D_{k+n}(0)$ of the initial state
at a point $\xi_1,\ldots,\xi_k;x_{k+1},\ldots,x_{k+n}$.

As a result of the formal transition to the thermodynamic limit,
Cauchy problem (\ref{z-Bs1})-(\ref{Bs2}) gets form (\ref{Gs1})-(\ref{Gs2}).

Let us transform equations (\ref{z-Bs1}) to the form which is usually used in the plasma theory.

We find the covariation of the microscopic phase density
$\langle\delta N^{(1)}(\xi_1) \delta N^{(1)}(\xi_2)\rangle$,
where the observable $\delta N^{(1)}= N^{(1)}-\langle N^{(1)}\rangle$
is a fluctuation of the microscopic phase density.

Since $ \langle \delta N^{(1)}(\xi_1) \delta N^{(1)}(\xi_2)\rangle=
\langle N^{(1)}(\xi_1) N^{(1)}(\xi_2)\rangle-\langle N^{(1)}(\xi_1) \rangle \langle N^{(1)}(\xi_2) \rangle $, we have
\begin{eqnarray*}
  && \langle N^{(1)}(\xi_1) N^{(1)}(\xi_2)\rangle(t)=\\
&&=\big(1,D(0)\big)^{-1}\sum\limits_{n=0}^{\infty}\frac{1}{n!}\int dx_{1}\ldots dx_{n}D_{n}(0,x_1,\ldots,x_n)\sum\limits_{i=1}^{n}\sum\limits_{j=1}^{n}\delta(\xi_1-X_{i}(t))\delta(\xi_2-X_{j}(t)).
\end{eqnarray*}
In accordance with the equality
\begin{eqnarray*}
 && \sum\limits_{i=1}^{n}\sum\limits_{j=1}^{n}\delta(\xi_1-x_{i})\delta(\xi_2-x_{j})=
 \sum\limits_{i=1}^{n}\delta(\xi_1-x_{i})\delta(\xi_2-x_{i})+
   \sum\limits_{i\neq j=1}^{n}\delta(\xi_1-x_{i})\delta(\xi_2-x_{j}),
\end{eqnarray*}
we find
\begin{eqnarray*}
  && \langle N^{(1)}(\xi_1) N^{(1)}(\xi_2)\rangle(t)
=\delta(\xi_1-\xi_2)\langle N^{(1)}\rangle(t,\xi_1)+\langle N^{(2)}\rangle(t,\xi_1,\xi_2).
\end{eqnarray*}

Thus, the covariation of the microscopic phase density $\langle\delta N^{(1)}(\xi_1) \delta N^{(1)}(\xi_2)\rangle$
is defined as follows:
\begin{eqnarray*}
  \langle\delta N^{(1)}(\xi_1) \delta N^{(1)}(\xi_2)\rangle(t)=\delta(\xi_1-\xi_2)\langle N^{(1)}\rangle(t,\xi_1)+
        \langle N^{(2)}\rangle(t,\xi_1,\xi_2)-\langle N^{(1)}\rangle(t,\xi_1)\langle N^{(1)}\rangle(t,\xi_2).
\end{eqnarray*}

For the regularized interaction potential, i.e. $\Phi^{'}(0)=0$, in terms of
the covariation of the microscopic phase density $\langle\delta N^{(1)}(\xi_1) \delta N^{(1)}(\xi_2)\rangle$,
the first equation from hierarchy (\ref{z-Bs1}) reduces to the Vlasov-type equation
\begin{eqnarray}\label{fin}
  && \frac{\partial}{\partial t}\langle N^{(1)}\rangle(t,\xi_1)+\langle v_1,
   \frac{\partial}{\partial r_1}\rangle\langle N^{(1)}\rangle(t,\xi_1)-\nonumber\\
  &&-\langle\frac{\partial}{\partial r_1}\int d\xi_2\Phi(r_1-r_2)\langle N^{(1)}\rangle(t,\xi_2),\frac{\partial}{\partial v_1}\rangle\langle N^{(1)}\rangle(t,\xi_1)=\nonumber\\
  &&=\int d\xi_2\langle\frac{\partial}{\partial r_1}\Phi(r_1-r_2),
     \frac{\partial}{\partial v_1}\rangle \langle\delta N^{(1)}(\xi_1) \delta N^{(1)}(\xi_2)\rangle(t).
\end{eqnarray}
This unclosed equation with respect to the average value of additive-type microscopic phase density $\langle N^{(1)}\rangle(t)$
is the conventional equation for the description of a system of charged particles \cite{Ba05}.

\subsection{The system of charged particles}
If we consider a plasma, i.e. a system of charged particles, then  $\Phi(r_1-r_2)= e^2\, |r_1-r_2|^{-1}$,
and the macroscopic electric field $\langle E\rangle(t)$ is defined from the equation
\begin{eqnarray*}
 && \mathrm{div}\langle E\rangle(t,\xi)= e\,\langle N^{(1)}\rangle(t,\xi)=\\
 &&= e\, \big(1,D(0)\big)^{-1}\sum\limits_{n=0}^{\infty}\frac{1}{n!}
  \int dx_{1}\ldots dx_{n}N^{(1)}_n(t,\xi)D_{n}(0),
\end{eqnarray*}
where $N^{(1)}_n(t,\xi)$ is the additive-type microscopic phase density (\ref{k-arN}).

Thus, in this case, equation (\ref{fin}) takes the form
\begin{eqnarray}\label{finK}
   &&\big(\frac{\partial}{\partial t}+ v^{\alpha}\frac{\partial}{\partial r^{\alpha}}+
   e\,\langle E^{\alpha}\rangle(t,r)\frac{\partial}{\partial v^{\alpha}}\big)\langle N^{(1)}\rangle(t,\xi)=
   -e\,\frac{\partial}{\partial v^{\alpha}}\langle\delta E^{\alpha}(t,r) \delta N^{(1)}(t,\xi)\rangle.
\end{eqnarray}
Here  $\delta E= E-\langle E\rangle$ is a fluctuation of the electric field $E=(E_1,\ldots,E_n,\ldots)$,
where
\begin{eqnarray*}
 && E_n\equiv E(t,r;x_1,\ldots,x_n)=
  -e\,\frac{\partial}{\partial r}\int d\xi'\, \frac{ N^{(1)}_n(t,\xi';x_1,\ldots,x_n)}{|r-r'|},
\end{eqnarray*}
and $\langle E\rangle\equiv\langle E\rangle(t,r)$ is the macroscopic electric field
\begin{eqnarray*}
&&\langle E\rangle(t,r)=-e\,\frac{\partial}{\partial r}\int d\xi' \frac{\langle N^{(1)}\rangle(t,\xi')}{|r-r'|}.
\end{eqnarray*}
Equation (\ref{finK}) is an unclosed equation with respect to the
average value of the microscopic phase density $\langle N^{(1)}\rangle(t)$.

\subsection{The hierarchy for average values of the microscopic phase densities:\\ the evolution of states}
According to (\ref{est}) and Cauchy problem (\ref{z-ks1})-(\ref{z-ks2}) of the Liouville equations the evolution equations
of the average values of the $k$-ary type microscopic phase densities get the form of hierarchy (\ref{z-Bs1})
with initial data (\ref{Bs2}).

Using the BBGKY hierarchy (\ref{1}), due to formula (\ref{avmarSch}) for marginal phase densities (\ref{ad_G0})  we derive
hierarchy (\ref{Gs1})  with initial data (\ref{Gs2}).

As we noted above, hierarchy (\ref{z-Bs1}) in the thermodynamic limit transforms to the hierarchy of equations (\ref{Gs1}).

\subsection{On solution of the initial-value problem}
To determine a solution of hierarchy (\ref{Gs1}), we introduce some preliminaries.

On integrable functions, we define the following group of operators
\begin{eqnarray}\label{eo}
&&\big(S_{k}(-t)f_{k}\big)(\xi_1,\ldots,\xi_k):=
f_{k}(\Xi_1(-t,\xi_1,\ldots,\xi_k),\ldots,\Xi_k(-t,\xi_1,\ldots,\xi_k)),
\end{eqnarray}
where the functions $\Xi_i(t)\equiv\big(V_i(t),R_i(t)\big)$  are the solution of the Cauchy problem of
the Hamilton equations for "macroscopic variables"\, (dynamics of continuum)
\begin{eqnarray}\label{he}
&&\frac{d }{d t}R_i(t)=V_i(t),\nonumber\\
&&\frac{d}{d t} V_i(t)=-\sum\limits_{j\neq i, j=1}^{k}\frac{\partial}{\partial R_{i}(t)}\Phi\big(R_i(t)-R_j(t)\big),
\end{eqnarray}
with the initial data
\begin{eqnarray*}
&&R_i(0)=r_i,\\
&&V_i(0)=v_i,\qquad i=1,\ldots, k.
\end{eqnarray*}
The generator of group (\ref{eo}) is defined by the Poisson bracket with respect to the variables $\xi_1,\ldots,\xi_k$
on the continuously differentiable functions $f_{k}\equiv f_{k}(\xi_1,\ldots,\xi_k)$
\begin{eqnarray*}
  &&\lim\limits_{t\rightarrow 0}\frac{1}{t}\big(S_{k}(-t)-I\big) f_{k}=
  \big(-\sum\limits_{i=1}^{k}\langle v_i,
  \frac{\partial}{\partial r_i}\rangle +
  \sum\limits_{i\neq j=1}^{k}\langle \frac{\partial}{\partial r_i}\Phi(r_i-r_j),
  \frac{\partial}{\partial v_i}\rangle \big)f_{k}.
\end{eqnarray*}

A solution of Cauchy problem (\ref{Gs1})-(\ref{Gs2}) is defined by the expansion
over the arity index of the microscopic phase density, whose evolution is governed by
the corresponding-order cumulant (semiinvariant) of the
evolution operators (\ref{eo}), namely
\begin{eqnarray}\label{RozvG}
&&\langle G^{(k)} \rangle (t,\xi_1,\ldots,\xi_k)=\big( U(t)\langle G \rangle (0)\big)_k(\xi_1,\ldots,\xi_k)=\nonumber\\
&&=\sum\limits_{n=0}^{\infty}\frac{1}{n!}\int d\xi_{k+1}\ldots d\xi_{k+n}
      \mathfrak{A}_{1+n}(-t,Y_{1},\xi_{k+1},\ldots,\xi_{k+n})\langle G^{(k+n)}\rangle (0),
\end{eqnarray}
where $\langle G\rangle(0)=(0,\langle G^{(1)}\rangle(0),\ldots,\langle G^{(k)}\rangle(0),\ldots)$
is the sequence of integrable functions. If $n \geq 0$,
\begin{eqnarray}\label{cm}
\mathfrak{A}_{1+n}(-t,Y_{1},\xi_{k+1},\ldots,\xi_{k+n})=
\sum\limits_{\mathrm{P}:\{Y_{1},X\setminus Y\} ={\bigcup\limits}_i X_i}(-1)^{|\mathrm{P}|-1}(|\mathrm{P}|-1)!
        \prod_{X_i\subset \mathrm{P}}S_{|X_i|}(-t,X_i)
\end{eqnarray}
is the $(1+n)th$-order cumulant of the groups of operators (\ref{eo}), ${\sum\limits}_\mathrm{P}$
is the sum over all possible partitions $\mathrm{P}$ of the set $\{Y_{1},\xi_{k+1},\ldots,\xi_{k+n}\}$ into
$|\mathrm{P}|$ nonempty mutually disjoint subsets $X_i\subset \{Y_{1},X\setminus Y\}\equiv\{Y_{1},\xi_{k+1},\ldots,\xi_{k+n}\}$.
The set $Y_1$ consists of one element of $Y\equiv(\xi_{1},\ldots,\xi_{k})$,
i.e. the set $(\xi_{1},\ldots,\xi_{k})$ is a connected subset of the partition $\mathrm{P}$ ($|\mathrm{P}|=1$).

If $\langle G^{(k)}\rangle (0)$ are integrable functions \cite{GerRS}, series (\ref{RozvG}) converges for small densities (see also \cite{CGP97},\cite{Co09}).

We remark that a solution of Cauchy problem (\ref{Gs1})-(\ref{Gs2}) can be represented as the iteration series
of the BBGKY hierarchy (\ref{Gs1})

\begin{eqnarray}\label{Iter}
     && \langle G^{(k)} \rangle (t,\xi_1,\ldots,\xi_k)= \sum\limits_{n=0}^{\infty} \int\limits_0^t dt_{1}\ldots\int\limits_0^{t_{n-1}}dt_{n}\,
          \int d\xi_{k+1}\ldots d\xi_{k+n}\times\nonumber\\
&&\times S_{k}(-t+t_{1})\sum\limits_{i_{1}=1}^{k}
          \big(-\mathcal{L}_{\mathrm{int}}(i_{1},k+1)\big)S_{k+1}(-t_{1}+t_{2})\ldots\nonumber\\
&&\ldots S_{k+n-1}(-t_{n-1}+t_{n})\sum\limits_{i_{n}=1}^{k+n-1}\big(-\mathcal{L}_{\mathrm{int}}(i_{n},k+n)\big)
    S_{k+n}(-t_{n})\langle G^{(k+n)}\rangle (0),
\end{eqnarray}
where
\begin{eqnarray*}
 && -\mathcal{L}_{\mathrm{int}}(i,j)\equiv\langle\frac{\partial}{\partial r_i}\Phi(r_i-r_j),
  \frac{\partial}{\partial v_i}\rangle.
\end{eqnarray*}
If we apply the Duhamel formula to cumulants of groups of operators (\ref{eo}), solution expansion (\ref{RozvG}) reduces
to iteration series (\ref{Iter}) of the BBGKY hierarchy.


\section{Evolution equations of non-Markovian transport}
Using above-obtained results, we develop a new approach to the description of the evolution of the
average value of microscopic phase density. Namely we formulate the non-Markovian evolution equation for
the average value of microscopic phase density.

\subsection{The generalized equation for an average value of the microscopic phase density}
Consider Cauchy problem (\ref{Gs1})-(\ref{Gs2}) with  initial data which completely defined by the average value
of additive-type microscopic phase density $\langle G^{(1)}\rangle(0)$, for example,
\begin{eqnarray}\label{fact}
        &&\langle G^{(k)}\rangle(0)=\prod_{i=1}^k \langle G^{(1)}\rangle(0,\xi_i).
\end{eqnarray}
We note that  $\langle G^{(1)}\rangle(0,\xi_i)=F_{1}(0,\xi_i)$. Therefore, initial data  (\ref{fact}) have the transparent sense:
 they satisfy the chaos condition.

In that case, the initial-value problem of the
BBGKY hierarchy (\ref{Gs1})-(\ref{Gs2}) is not a completely well-defined Cauchy problem,
because the generic initial data are not independent for every
unknown functions $\langle G^{(k)}\rangle(t),\, k\geq 1$, of the
hierarchy of equations. Thus, it naturally arises the possibility of
reformulating such initial-value problem as a new Cauchy problem for
independent unknown function, i.e. the average value
of the additive-type microscopic phase density $\langle G^{(1)}\rangle(t)$, together with
explicitly defined functionals $ \langle G^{(k)}\rangle\big(t,\xi_1,\ldots,\xi_k \mid \langle G^{(1)}\rangle(t)\big),\, k\geq 2,$
of the solution $\langle G^{(1)}\rangle(t)$
of this Cauchy problem instead other unknown average values
of microscopic phase densities  \cite{CGP97}, \cite{GP98}.

The functionals $\langle G^{(k)}\rangle \big(t,\xi_1,\ldots,\xi_k \mid \langle G^{(1)}\rangle(t)\big),\, k\geq2,$
are represented by the expansions
over the products with respect to the average value of the additive-type microscopic phase density $\langle G^{(1)}\rangle(t)$
\begin{eqnarray}\label{f}
   \langle G^{(k)}\rangle\big(t,\xi_1,\ldots,\xi_k \mid \langle G^{(1)}\rangle(t)\big):=
  \sum\limits_{n=0}^{\infty }\frac{1}{n!}\int d\xi_{k+1}\ldots d\xi_{k+n} \,
           \mathfrak{V}_{1+n}(t)\prod _{i=1}^{k+n} \langle G^{(1)}\rangle(t,\xi_i),
\end{eqnarray}
where the evolution operators $\mathfrak{V}_{1+n}(t)\equiv\mathfrak{V}_{1+n}(t,Y_1,\xi_{k+1},\ldots, \xi_{k+n})$ are defined
from the condition that functionals $\langle G^{(k)}\rangle(t)\big(t \mid \langle G^{(1)}\rangle(t)\big)$
are congruent with solutions (\ref{RozvG}) of initial-value problem (\ref{Gs1})-(\ref{Gs2}).

We give examples of first terms of series (\ref{f}):
\begin{eqnarray*}
 && \mathfrak{V}_{1}(t,Y_{1})=\widehat{\mathfrak{A}}_{1}(t,Y_{1}),\\
&&\mathfrak{V}_{2}(t,Y_{1},\xi_{k+1})=
\widehat{\mathfrak{A}}_{2}(t,Y_{1},\xi_{k+1})-\widehat{\mathfrak{A}}_{1}(t,Y_{1})\sum_{j=1}^s
 \widehat{\mathfrak{A}}_{2}(t,\xi_{j},\xi_{k+1}),
\end{eqnarray*}
where $Y_1\equiv(\xi_1,\ldots,\xi_k)$ and $\widehat{\mathfrak{A}}_{k}(t)$ is the $kth$-order, $k\geq1$, cumulant of
the scattering operators
\begin{eqnarray}\label{so}
 &&\widehat{S}_{k}(t)=S_{k}(-t,\xi_1,\ldots,\xi_k)\prod\limits_{i=1}^{k}S_{1}(t,\xi_{i}),
\end{eqnarray}
the operator $S_{k}(-t,\xi_1,\ldots,\xi_k)$ is defined by formula (\ref{eo}), and $\widehat{S}_{1}(t)=I$ is the identity operator.

On integrable functions, the action of scattering operators (\ref{so}) is defined by the formula
\begin{eqnarray}\label{sod}
\big(\widehat{S}_{k}(t)f_{k}\big)(\xi_1,\ldots,\xi_k)=
  f_{k}\Big(\Xi_1\big(t,\,\Xi_1(-t,\xi_1,\ldots,\xi_k)\big),\ldots
   ,\Xi_k\big(t,\,\Xi_k(-t,\xi_1,\ldots,\xi_k)\big)\Big),
\end{eqnarray}
where the functions $\Xi_i(t)\equiv\big(V_i(t),R_i(t)\big)$,\,$i=1,\ldots,n$  are solutions of the Cauchy problem of
Hamilton equations (\ref{he}) for "macroscopic variables"\, with corresponding initial data.
The generator of group (\ref{so}) of scattering operators is defined by the Poisson bracket
with respect to the variables $\xi_1,\ldots,\xi_k$
with an interaction potential $\Phi$ on the continuously differentiable functions $f_{k}\equiv f_{k}(\xi_1,\ldots,\xi_k)$
\begin{eqnarray*}
  &&\lim\limits_{t\rightarrow 0}\frac{1}{t}\big(\widehat{S}_{k}(t)-I\big) f_{k}=
  \sum\limits_{i\neq j=1}^{k}\langle \frac{\partial}{\partial r_i}\Phi(r_i-r_j),
  \frac{\partial}{\partial v_i}\rangle f_{k}.
\end{eqnarray*}

In terms of scattering operators (\ref{so}), the first terms of series (\ref{f})
have the form
\begin{eqnarray*}
&&\mathfrak{V}_{1}(t,Y_{1})=\widehat{S}_{k}(t,Y),\\
&&\mathfrak{V}_{2}(t,Y_{1},\xi_{k+1})= \widehat{S}_{k+1}(t,Y,\xi_{k+1})-
  \widehat{S}_k(t,Y)\sum_{j=1}^k  \widehat{S}_2(t,\xi_{j},\xi_{k+1})+(k-1)\widehat{S}_k(t,Y).
\end{eqnarray*}

If $\langle G^{(1)}\rangle(t)$ is an integrable function,
then, in case of low densities, expansion (\ref{f}) is the converging series \cite{GP98}.

The average value of the additive-type microscopic phase density $\langle G^{(1)}\rangle(t)$ is governed by
the following Cauchy problem (\emph{the generalized evolution
equation for the average value of microscopic phase density}):
\begin{eqnarray}\label{gke}
&&\frac{\partial}{\partial t}\langle G^{(1)}\rangle(t,\xi_1)+\langle v_1,\frac{\partial}{\partial r_1}\rangle\langle G^{(1)}\rangle(t,\xi_1)=\nonumber\\
  && =\int d\xi_{2}\langle\frac{\partial}{\partial r_1}\Phi(r_1-r_2),\frac{\partial}{\partial v_1}\rangle
\langle G^{(2)}\rangle\big(t,\xi_1,\xi_2 \mid \langle G^{(1)}\rangle(t)\big)
\end{eqnarray}
with the initial data
\begin{eqnarray}\label{gke2}
         &&  \langle G^{(1)}\rangle(t,\xi_1)|_{t=0}= \langle G^{(1)}\rangle(0,\xi_1).
\end{eqnarray}
The functional $\langle G^{(2)}\rangle\big(t,\xi_1,\xi_2 \mid \langle G^{(1)}\rangle(t)\big)$ in the collision integral
of equation (\ref{gke}) is defined by expansion (\ref{f})
\begin{eqnarray}\label{ff}
&&\langle G^{(2)}\rangle(t)\big(t,\xi_1,\xi_2 \mid \langle G^{(1)}\rangle(t)\big)=
   \widehat{\mathfrak{A}}_{1}(t,Y_{1})\prod_{i=1}^2 \langle G^{(1)}\rangle(t,\xi_i)+\nonumber\\
&&+\int d\xi_{3}\big(\widehat{\mathfrak{A}}_{2}(t,Y_{1},\xi_{k+1})-
 \widehat{\mathfrak{A}}_{1}(t,Y_{1})\sum_{j=1}^s \widehat{\mathfrak{A}}_{2}(t,\xi_{j},\xi_{k+1}) \big)
 \prod_{i=1}^{3} \langle G^{(1)}\rangle(t,\xi_i) + ...,
\end{eqnarray}
where $Y_1\equiv(\xi_1,\xi_2)$.

We represent the first term of expansion (\ref{ff}) in an explicit form.
According to the definition of two-particle scattering operator (\ref{sod}), we have
\begin{eqnarray}\label{ffff}
&& \widehat{\mathfrak{A}}_{1}(t,Y_{1})\prod_{i=1}^2 \langle G^{(1)}\rangle(t,\xi_i)=\\
&&=\langle G^{(1)}\rangle \big(t,\,\Xi_1\big(t,\,\Xi_1(-t,\,\xi_1,\xi_2)\big)\big)\langle G^{(1)}\rangle \big(t,\,\Xi_2\big(t,\,\Xi_2(-t,\,\xi_1,\xi_2)\big)\big),\nonumber
\end{eqnarray}
where the functions ($i=1,2$)
\begin{eqnarray*}
 && \Xi_i\big(t,\,\Xi_i(-t,\,\xi_1,\xi_2)\big)=
\big(V_i(-t,\,\xi_1,\xi_2),R_i(-t,\,\xi_1,\xi_2)+t\,V_i(-t,\,\xi_1,\xi_2)\big)
\end{eqnarray*}
are defined as above by (\ref{he}) in formula (\ref{sod}).

The following statement is true.
If the initial data are completely defined by $\langle G^{(1)}\rangle(0)$, then Cauchy problem (\ref{Gs1})-(\ref{Gs2})
is equivalent to initial-value problem (\ref{gke})-(\ref{gke2})
of the generalized evolution equation for the average value of the microscopic phase density
and the functionals $\langle G^{(k)}\rangle\big(t,\xi_1,\ldots,\xi_k \mid \langle G^{(1)}\rangle(t)\big)$, \,$k\geq2$, defined by
expansions (\ref{f}) provided a low density of particles.

Thus, at an arbitrary moment time
the evolution of the average value of the microscopic phase density can be described
by equation (\ref{gke}) without any approximations.

The solution of Cauchy problem (\ref{gke})-(\ref{gke2}) is defined by the expansion
\begin{eqnarray}\label{Rozvgke}
&&\langle G^{(1)} \rangle (t,\xi_1)=
\sum\limits_{n=0}^{\infty}\frac{1}{n!}\int d\xi_{2}\ldots d\xi_{1+n}\,
      \mathfrak{A}_{1+n}(-t,\xi_{1},\ldots,\xi_{n+1})\prod_{i=1}^{n+1} \langle G^{(1)}\rangle(0,\xi_i),
\end{eqnarray}
where $\mathfrak{A}_{1+n}(-t,\xi_{1},\ldots,\xi_{n+1})$
is the $(1+n)th$-order cumulant of the group of operators (\ref{eo}).

For the low densities, i.e. $\int |\langle G^{(1)}\rangle(0,\xi)|d\xi<e^{-1}$, series (\ref{Rozvgke}) converges.
If $\langle G^{(1)}\rangle(0)$ is a continuously differentiable function with compact support, then, for $t\in \mathbb{R}$,
it is a classical solution of generalized equation (\ref{gke}) for the average value of the microscopic phase density.

We observe the links of the introduced generalized equation for the average value of the microscopic phase density
with Bogolyubov's method of the derivation of kinetic equations \cite{BC}.
Functionals (\ref{f}) are formally similar to the corresponding functionals
of Bogolyubov's method \cite{Co09} if they satisfy the principle of weakening of correlations \cite{CGP97}.
The proof of this statement is completely similar to the proof of the equivalence of both
representations of the BBGKY hierarchy solutions as iteration series (\ref{Iter}) and functional series (\ref{RozvG}).

\subsection{An example: the regularized interaction potential}
We separate the Vlasov term in equation (\ref{gke}) to represent it in the conventional form
for the description of a system of charged particles.

We introduce the macroscopic characteristic
of correlations in continuum
\begin{eqnarray}\label{ec}
 \langle \mathcal{G}\rangle\big(t,\xi_1,\xi_2\mid \langle G^{(1)}\rangle(t)\big)=
  \langle G^{(2)}\rangle(t)\big(t,\xi_1,\xi_2 \mid \langle G^{(1)}\rangle(t)\big)-
  \langle G^{(1)}\rangle(t,\xi_1)\langle G^{(1)}\rangle(t,\xi_2),
\end{eqnarray}
where the functional  $\langle G^{(2)}\rangle(t)\big(t,\xi_1,\xi_2 \mid \langle G^{(1)}\rangle(t)\big)$ is defined
by expansion (\ref{ff}). Then the generalized evolution equation (\ref{gke})
for the average value of microscopic phase density gets the form

\begin{eqnarray}\label{final}
&&\frac{\partial}{\partial t}\langle G^{(1)}\rangle(t,\xi_1)=-\langle v_1,\frac{\partial}{\partial r_1}\rangle+
   \langle\frac{\partial}{\partial r_1}\int d\xi_2\Phi(r_1-r_2)\langle G^{(1)}\rangle(t,\xi_2),
     \frac{\partial}{\partial v_1}\rangle \langle G^{(1)}\rangle(t,\xi_1)+\nonumber\\
&&+\int d\xi_2 \langle\frac{\partial}{\partial r_1}\Phi(r_1-r_2),\frac{\partial}{\partial v_1}\rangle
    \langle \mathcal{G}\rangle\big(t,\xi_1,\xi_2\mid \langle G^{(1)}\rangle(t)\big)
\end{eqnarray}
with initial data (\ref{gke2}).

In the case of regularized interaction potential, i.e. $\Phi^{'}(0)=0$, the functional
$\langle \mathcal{G}\rangle\big(t,\xi_1,\xi_2\mid \langle G^{(1)}\rangle(t)\big)$ is the formal thermodynamic
limit of the covariation of the microscopic phase density $\langle\delta N^{(1)}(\xi_1) \delta N^{(1)}(\xi_2)\rangle$
from the right hand side of equation (\ref{fin}).

If we consider a plasma, i.e. a system of charged particles,
the macroscopic electric field $\langle E\rangle(t)$ is determined from the equation
\begin{eqnarray*}
  &&\mathrm{div}\langle E\rangle(t,\xi)= e\,\langle G^{(1)}\rangle(t,\xi).
\end{eqnarray*}
In this case, equation (\ref{final}) gets the form
\begin{eqnarray*}
  \big(\frac{\partial}{\partial t}+ v^{\alpha}\frac{\partial}{\partial r^{\alpha}}+
   e\,\langle E^{\alpha}\rangle(t,r)\frac{\partial}{\partial v^{\alpha}}\big)\langle G^{(1)}\rangle(t,\xi)=\int d\xi' \,\frac{\partial}{\partial r^{\alpha}}\,\frac{1}{|r-r'|}\, \frac{\partial}{\partial v^{\alpha}}
    \,\langle \mathcal{G}\rangle\big(t,\xi,\xi'\mid \langle G^{(1)}\rangle(t)\big),
\end{eqnarray*}
where the functional $\langle\mathcal{G}\rangle\big(t\mid \langle G^{(1)}\rangle(t)\big)$ in the collision integral
is defined by expansions (\ref{ec}),(\ref{ff}).

\subsection{On the Fokker-Planck representation of a generalized collision integral}
If we consider a plasma \cite{BC}, then the small parameter is the plasma parameter associated with the weak interaction
\begin{eqnarray*}
  \varepsilon=\beta e^2/r_D,
\end{eqnarray*}
where $r_D$ is the Debye screening parameter, $e$ is the electric charge and $\beta^{-1}$
is the value of the order of the average kinetic energy of electrons.

Let us consider a first term of the collision integral expansion of
$\langle \mathcal{G}\rangle\big(t,\xi_1,\xi_2\mid \langle G^{(1)}\rangle(t)\big)$ in equation (\ref{final}),
i.e. the first order term with respect to the plasma parameter associated with the weak interaction,
\begin{eqnarray}\label{apI}
   &&\mathcal{I}\equiv
   \int d\xi_2\big\langle\frac{\partial}{\partial r_1}\Phi(r_1-r_2),\frac{\partial}{\partial v_1}\big\rangle
   \big(\widehat{S}_2(t,\xi_1,\xi_2)-
   I\big) \prod_{i=1}^2 \langle G^{(1)}\rangle(t,\xi_i).
\end{eqnarray}

We will construct a suitable approximation of the non-Markovian collision integral (\ref{apI}) for a plasma.
Due to the fact that, for the scattering operator $\widehat{S}_2(t,1,2)$, the Duhamel formula is formally true
(see also (\ref{ffff}))
\begin{eqnarray*}
&&\widehat{S}_2(t,\xi_1,\xi_2)-I =\int\limits_{0}^{t}d\tau S_2(-t+\tau,\xi_1,\xi_2)
   \big(\langle \frac{\partial}{\partial r_1}\Phi(r_1-r_2),\frac{\partial}{\partial v_1} \rangle+\\
&&+\langle\frac{\partial}{\partial r_2}\Phi(r_1-r_2),\frac{\partial}{\partial v_2} \rangle \big)S_1(-\tau+t,\xi_1)S_1(-\tau+t,\xi_2),
\end{eqnarray*}
where $S_2(-t+\tau,\xi_1,\xi_2)$ is evolution operator (\ref{eo}),
collision integral (\ref{apI}) can be given as follows:
\begin{eqnarray*}
   &&\mathcal{I}=\int\limits_{0}^{t}d\tau\int d\xi_2 \langle\frac{\partial}{\partial r_1}\Phi(r_1-r_2),\frac{\partial}{\partial v_1} \rangle S_2(-t+\tau,\xi_1,\xi_2)
   \big(\langle\frac{\partial}{\partial r_1}\Phi(r_1-r_2),\frac{\partial}{\partial v_1} \rangle +\\
   &&+\langle\frac{\partial}{\partial r_2}\Phi(r_1-r_2),\frac{\partial}{\partial v_2} \rangle \big) S_1(-\tau+t,\xi_1)S_1(-\tau+t,\xi_2)\langle G^{(1)}\rangle(t,\xi_1)\langle G^{(1)}\rangle(t,\xi_2).
\end{eqnarray*}

Then, by expanding the obtained collision integral $\mathcal{I}$ in the Dyson$-$Phillips series
in the plasma parameter and by restricting ourselves to the first term of the expansion (weak coupling limit), we find
\begin{eqnarray*}
  && \mathcal{I}=\int\limits_{0}^{t} d\tau \int d\xi_2 \langle\frac{\partial}{\partial r_1}\Phi(r_1-r_2),
   \frac{\partial}{\partial v_1} \rangle \langle\frac{\partial}{\partial r_1}\Phi\big(r_1(-t+\tau)-\\
   &&-r_2(-t+\tau)\big),
  \big(\frac{\partial}{\partial v_1}- \frac{\partial}{\partial v_2}\big)\rangle \langle G^{(1)}\rangle(t,\xi_1)\langle G^{(1)}\rangle(t,\xi_2),
\end{eqnarray*}
where $r_i(-t+\tau)\equiv r_i(-t+\tau,\xi_1,\xi_2)$,\, $i=1,2$, are the solutions
of initial-value problem (\ref{he}) for $k=2$.
We remark that this expression has similar structure to the generalized Landau collision integral
in the kinetic theory \cite{BC}.

Thus, we can finally rewrite equation (\ref{final})
in the Fokker-Planck representation \cite{GHT} as
\begin{eqnarray}\label{FP}
 && \frac{\partial}{\partial t}\langle G^{(1)}\rangle(t,\xi_1)+\langle v_1,\frac{\partial}{\partial r_1} \rangle \langle G^{(1)}\rangle(t,\xi_1)-\nonumber\\
 &&- \langle\frac{\partial}{\partial r_1}\int d\xi_2\Phi(r_1-r_2)\langle G^{(1)}\rangle(t,\xi_2),
     \frac{\partial}{\partial v_1}\rangle \langle G^{(1)}\rangle(t,\xi_1)=\nonumber\\
&&=\frac{\partial}{\partial v^{\alpha}_1 \partial v^{\beta}_1}\mathcal{D}^{\alpha\beta}(t)\langle G^{(1)}\rangle(t,\xi_1)+
  \frac{\partial}{\partial v^{\alpha}_1}\mathcal{B}^{\alpha}(t)\langle G^{(1)}\rangle(t,\xi_1),
\end{eqnarray}
where the transport coefficients are defined by the expressions
\begin{eqnarray}\label{dc}
   &&\mathcal{D}^{\alpha\beta}(t)= \int\limits_{0}^{t}d\tau\int d\xi_2
   \frac{\partial}{\partial r^{\alpha}_1}\Phi(r_1-r_2)\frac{\partial}{\partial r^{\beta}_1}\Phi\big(r_1(-t+\tau)-r_2(-t+\tau)\big)\langle G^{(1)}\rangle(t,\xi_2),\\
\label{fc}
  && \mathcal{B}^{\alpha}(t)=
  \int\limits_{0}^{t}d\tau\int d\xi_2\frac{\partial}{\partial r^{\alpha}_1}\Phi(r_1-r_2)\Big(\frac{\partial}{\partial r_2^{\beta}}\Phi\big(r_1(-t+\tau)-r_2(-t+\tau)\big)\frac{\partial}{\partial v_2^{\beta}}-\nonumber\\
&&-\frac{\partial}{\partial v_1^{\beta}}\frac{\partial}{\partial r_1^{\beta}}\Phi\big(r_1(-t+\tau)-
  r_2(-t+\tau)\big)\Big)\langle G^{(1)}\rangle(t,\xi_2).
\end{eqnarray}

Evolution equation (\ref{FP}) is a non-Markovian evolution equation for the average value of the microscopic phase density with
the Fokker-Planck collision integral determining by nonlinear transport coefficients (\ref{dc}),(\ref{fc}). It can be reduced to
the canonical  Fokker-Planck equation as a result of further approximations \cite{Ba05}.

We observe that the higher-order terms in (\ref{final}) with respect to the plasma parameter from the functional
$\langle \mathcal{G}\rangle\big(t,\xi_1,\xi_2\mid \langle G^{(1)}\rangle(t)\big)$ are equal to zero
in the weak coupling approximation.

\section{Conclusion}
The BBGKY hierarchy of equations (\ref{Gs1}) for the generalized microscopic phase densities is formulated, and
the microscopic derivation of a non-Markovian collision term in the evolution equation
for the average value of the additive-type microscopic phase density (\ref{gke}) is done.

The time-nonlocal generalization of the Fokker-Planck equation (\ref{FP})-(\ref{fc})
on the basis of such collision term (\ref{apI}) is proposed.
The evolution equation generated by such collision term can be applied for
further analysis of turbulent transport \cite{Ba05}. The memory effects can be important for the description of
transport under saturated turbulence \cite{ZW99,ZW01,ZW09}.


\end{document}